\documentclass[journal]{IEEEtran}
\usepackage{blindtext}
\usepackage{graphicx}
\usepackage{amssymb}
\usepackage{amsmath}
\usepackage{amsthm}
\usepackage[english]{babel}
\usepackage{xspace}
\usepackage{cases}
\usepackage{color}
\usepackage{graphicx}
\usepackage[caption=false]{subfig}
\usepackage{epstopdf}
\usepackage{upgreek}
\usepackage{booktabs}
\usepackage{array}
\usepackage{multirow}
\usepackage{bbold}
\usepackage{soul}
\usepackage{enumerate}
\usepackage{csquotes}

\newtheorem{ass}{Assumption}

\newtheorem{exmpl}{Example}

\newcommand{\unaryminus}{\scalebox{0.5}[1.0]{\( - \)}}

\newcommand{\nbus}{N}

\newcommand{\nomega}{N_\omega}

\newcommand{\rsd}{s}

\newcommand{\iplain}{i}
\newcommand{\iRV}{\rv{\iplain}}

\newcommand{\vplain}{v}
\newcommand{\vre}{\vplain^{\text{re}}}
\newcommand{\vreRV}{\rv{\vplain}^{\text{re}}}
\newcommand{\vRV}{\rv{\vplain}}

\newcommand{\vim}{\vplain^{\text{im}}}%
\newcommand{\vimRV}{\rv{\vplain}^{\text{im}}}

\newcommand{\pplain}{p}
\newcommand{\qplain}{q}
\newcommand{\p}{\pplain^{\text{\normalfont g}}}
\newcommand{\punc}{\pplain^{\text{\normalfont u}}}
\newcommand{\q}{\qplain^{\text{\normalfont g}}}
\newcommand{\qunc}{\qplain^{\text{\normalfont u}}}
\newcommand{\pd}{\pplain^{\text{\normalfont d}}}
\newcommand{\pdnom}{\pplain^{\text{\normalfont d,\,nom}}}

\newcommand{\pdRV}{\rv{\pplain}^{\text{\normalfont d}}}

\newcommand{\pRV}{\rv{\pplain}^{\text{\normalfont g}}}
\newcommand{\qRV}{\rv{\qplain}^{\text{\normalfont g}}}
\newcommand{\puncRV}{\rv{\pplain}^{\text{\normalfont u}}}
\newcommand{\quncRV}{\rv{\qplain}^{\text{\normalfont u}}}

\newcommand{\basisfun}{\psi}

\newcommand{\adjustlength}{-4.8mm}
\newcommand{\adjustOPF}{-10mm}
\newcommand{\adjustOPFmod}{-14mm}
\newcommand{\tablecaptionsep}{-2mm}
\newcommand{\rowstretch}{0.70}
\newcommand{\opf}{\textsc{opf}\xspace}

\newcommand{\dc}{\textsc{dc}\xspace}
\newcommand{\ac}{\textsc{ac}\xspace}
\newcommand{\acopf}{\textsc{ac}-\textsc{opf}\xspace}
\newcommand{\dcopf}{\textsc{dc}-\textsc{opf}\xspace}

\newcommand{\pce}{\textsc{pce}\xspace}
\newcommand{\pces}{{\textsc{pce}}s\xspace}

\newcommand{\cc}{\textsc{cc}}

\newcommand{\ccopf}{\cc-\opf}

\newcommand{\rv}[1]{\mathsf{#1}}

\newcommand{\ev}[1]{\mathbb{E}[#1]}
\newcommand{\evbig}[1]{\mathbb{E}\Big[#1\Big]}
\newcommand{\var}[1]{\mathbb{V}[#1]}

\newcommand{\pvar}{u}
\newcommand{\pvarup}{\expandafter\uppercase\expandafter{\pvar}}

\newcommand{\pfix}{d}
\newcommand{\pfixup}{\expandafter\uppercase\expandafter{\pfix}}

\definecolor{till}{rgb}{.1,.4,.9}
\definecolor{timm}{rgb}{.30,.60,.30} 
\definecolor{sidhant}{rgb}{.8,.1,.1} 
\definecolor{line}{rgb}{0.4,.4,.4}
\definecolor{veit}{rgb}{0.,.8,.4}
\definecolor{revised}{rgb}{.2,.8,.1}

\def\till{\textcolor{till}}

\def\sidhant{\textcolor{sidhant}}

\usepackage[bibstyle=numeric,citestyle=numeric-comp, uniquename=false, maxbibnames=6, maxcitenames=2, doi=false, url=false, eprint=false, isbn=false, backend=bibtex, sorting=none]{biblatex}

\bibliography{lit.bib}
\usepackage[table]{xcolor}
\ifCLASSINFOpdf
\else
\fi
\newcommand{\PaperTitle}{Chance-Constrained AC Optimal Power Flow \\-- A Polynomial Chaos Approach}
\usepackage{siunitx}

\let\labelindent\relax
\begin{document}
	%
	\title{\PaperTitle}
	%
	%
	%
	
	\author{Tillmann~M\"uhlpfordt,$^\text{a}$~Line~Roald,$^\text{b}$~Veit~Hagenmeyer,$^\text{a}$~Timm~Faulwasser,$^\text{a}$~Sidhant~Misra$^\text{c}$
		\thanks{\textsuperscript{a}Institute for Automation and Applied Informatics (\textsc{iai}), Karlsruhe Institute of Technology (\textsc{kit}), Karlsruhe, Germany, $\{$tillmann.muehlpfordt, veit.hagenmeyer$\}$@kit.edu, timm.faulwasser@ieee.org.}
		\thanks{\textsuperscript{b}Electrical and Computer Engineering, University of Wisconsin - Madison, Madison, \textsc{wi}, \textsc{usa}, roald@wisc.edu.}
		\thanks{\textsuperscript{c}Los Alamos National Laboratory,\,Los Alamos,\,\textsc{nm},\,\textsc{usa}, sidhant@lanl.gov.}
}

	\maketitle

	\begin{abstract}
	    As the share of renewables in the  grid increases, the operation of power systems becomes more challenging.
	    The present paper proposes a method to formulate and solve chance-constrained optimal power flow while explicitly considering the full nonlinear \ac power flow equations and stochastic uncertainties.
	    We use polynomial chaos expansion to model the effects of arbitrary uncertainties of finite variance, which enables to predict and optimize the system state for a range of operating conditions.
	    We apply chance constraints to limit the probability of violations of inequality constraints. 
		Our method incorporates a more detailed and a more flexible description of both the controllable variables and the resulting system state than previous methods.
		Two case studies highlight the efficacy of the method, with a focus on satisfaction of the \ac power flow equations and on the accurate computation of moments of all random variables.
	\end{abstract}
	
	\begin{IEEEkeywords}
	AC optimal power flow, uncertainty, polynomial chaos expansion, chance constraints
	\end{IEEEkeywords}

	%
	\IEEEpeerreviewmaketitle
	
	
	\ifCLASSOPTIONcaptionsoff
	\newpage
	\fi
	\section{Introduction}
	\label{sec:Introduction}
	The share of electricity generated through renewable energy sources such as wind and solar is increasing across the world~\cite{REN21Report}.
	This trend renders the operation of power systems more uncertain and more variable, which in turn has implications on all aspects of power systems operation---prompting a need for new and improved methods for uncertainty-aware scheduling and control.
    The present paper addresses the issue of modelling and mitigating the impact of uncertainty in optimal power flow (\opf) problems.
    These problems constitute an essential building block in power system operational processes such as market clearing \cite{stott2012} or security assessment \cite{capitanescu2011stateoftheart}.
    Much of the existing literature has focused on modelling and solving stochastic optimization problems based on the \dc power flow equations, including chance-constrained and robust versions of \dcopf \cite{Vrakopoulou12, Roald13, Bienstock14, Warrington13, Muehlpfordt18b}.
    The increasing interest in using stochastic \opf in applications that require more detailed modelling of reactive power and voltage magnitudes, such as voltage control or distribution grid optimization, has inspired the development of methods that incorporate the full nonlinear \ac power flow equations.
    Existing approaches include, e.g., chance constraints \cite{vrakopoulou2013AC, Zhang10, Zhang11, venzke2018, Roald18, dallanese2017}, robust formulations \cite{lorca2017robust, nasri2016, louca2017, lorca2017robust, molzahn18} \opf, and distributionally robust approaches \cite{Duan18}.
    The handling of the \ac power flow constraints under stochastic nodal injections  is notoriously difficult, because (i) it requires propagating uncertainties through a set of implicit nonlinear equations; 
    and (ii) algorithms that provide probabilistic or robust guarantees for constraint satisfaction often exploit convexity of the underlying optimization problem, which is not true for \acopf.
    The present paper addresses both of the above issues.
    
    Existing methods have coped with these challenges differently.
    For example, \cite{vrakopoulou2013AC, venzke2018, nasri2016, lorca2017robust} use convex relaxations of the power flow constraints, allowing to use established stochastic optimization algorithms.
    Since the relaxations expand the feasible space, the solutions are however not guaranteed to be feasible for the original chance-constrained or robust \acopf problem.
    In \cite{molzahn18}, convex relaxations are used to provide a conservative estimate of the uncertainty impact, which guarantees robust constraint satisfaction, but sacrifices performance of the solution.
    The authors of \cite{louca2017} propose a robust \acopf based on convex inner approximations, however, the method requires controllable power injections at every node as it cannot satisfy the nodal power balance constraints with equality.
    Related approaches use a full or partial linearization of the \ac power flow equations \cite{dallanese2017, Roald18},
    where the \ac power flow constraints are linearized around an operating point and the problem is solved using methods similar to the approaches developed for the convex \dcopf problem.
    
    In the present paper we propose a tractable formulation of chance-constrained \acopf that essentially satisfies the full nonlinear \ac power flow equations for generic uncertainties of finite variance, without relying on samples, relaxations or linearizations.
    Our problem formulation is based on polynomial chaos expansion (\pce), a spectral method for random variables analogous to a ``Fourier series for random variables'' \cite{Xiu10book}.
    Polynomial chaos allows to propagate uncertainty from the inputs to the relevant quantities such as current flows and voltage magnitudes, while accounting for the full nonlinearity of \ac power flow.
    The accuracy of \pce-based problem formulations as well as the computational tractability depends on the maximum degree of the underlying polynomial basis.
    Theoretically, an infinite degree is required to satisfy the \ac power flow equations \emph{exactly}.
    However, we show by means of experiments that the \ac power flow equations can be satisfied to high numerical accuracy for all uncertainty realizations already with low maximum degrees of about two or three.
    Furthermore, \pce facilitates moment-based reformulations of chance constraints, since the moments of all random variables can be computed directly from the \pce representation.
    In contrast to existing methods, \pce requires no linearization or sampling for either the moment computation or the chance-constrained formulation.
   
    
    The primary advantage of \pce for stochastic \opf is its ability to accurately and efficiently handle \emph{equality} constraints that involve random variables such as the full nonlinear \ac power flow equations under uncertainty.
    At the same time \pce also helps enforcing \emph{inequality} constraints using moment-based reformulations of chance constraints.
    While several alternatives have been explored in the literature to enforce inequality constraints under uncertainty---such as distributionally robust formulations where the uncertainty is modelled by a family of distributions that have matching (first two) moments~\cite{Zymler2013})---structured methods that enforce equality constraints involving random variables remain less studied. For example,
    \cite{Roald18, Duan18} formulate chance-constrained and distributionally robust chance-constrained versions of the \acopf, but the \ac power flow equations are satisfied only for the expected value while deviations are modelled through a linearization.
    Polynomial chaos allows a more elegant approach to \ac  power flow. 

    Polynomial chaos has been applied previously to power system optimization.
    For stochastic \opf under the \dc approximation it has been  shown that \pce provides exact and tractable convex reformulations~\cite{Muehlpfordt17a,Muehlpfordt18b}.
    With \ac equations, \cite{Muehlpfordt16b, Engelmann18b} apply polynomial chaos to formulate the problem.
    However, \cite{Muehlpfordt16b} considers only constraints for the expected values of generated powers, and \cite{Engelmann18b} does neither account for voltage magnitude constraints nor for line limits.
    Recently,  \pce has been applied to the multi-period \acopf problem under uncertainty in~\cite{Li18}; a conic relaxation of the power flow equations is employed together with sparse regression to compute the \pce coefficients, based on the method from~\cite{Muehlpfordt16b}.
    All works \cite{Muehlpfordt16b,Engelmann18b,Li18} lack a thorough probabilistic analysis of the satisfaction of the \ac power flow equations, as well as a validation of the moments (mean and variance) of the power system state variables, such as line currents and bus voltages magnitudes. 
    The present paper aims to close that gap.  The contributions are as follows:
   (i) a framework to formulate chance-constrained \acopf using \pce as a one-shot optimization problem, accounting for voltage magnitude and current magnitude limits, but without relying on samples, relaxations or linearizations;
        (ii) investigation of \ac power flow satisfaction for varying maximum degrees of the \pce basis; and
        (iii) validation of accuracy of moments, and comparison to linearized \ac power flow.
        (iv) validation of empirical constraint satisfaction via in- and out-of-sample tests.

    \emph{Paper organization:} Section~\ref{sec:ProblemFormulation} discusses the power system model, the uncertainty model, and the chance-constrained \opf problem.
    Section~\ref{sec:PCE} introduces \pce and its advantages for \acopf.
    Section~\ref{sec:CCOPF_UsingPCE} applies \pce to chance-constrained \opf, and provides a tractable reformulation.
    The case studies (for a 5- and 30-bus system) from Section~\ref{sec:SimulationStudies} demonstrate the efficacy of the proposed approach.

	\section{Problem Formulation}
	\label{sec:ProblemFormulation}
	\subsection{Power System Model}
	\label{sec:PowerSystemModel}
	Consider a connected $\nbus$-bus electrical network represented by its set of bus indices $\mathcal{N} = \{1, \hdots, \nbus \}$, and its set of line indices $\mathcal{L} \subseteq \mathcal{\nbus} \times \mathcal{\nbus}$.
	At each bus $i \in \mathcal{\nbus}$, we define the complex power $s_i = \pplain_i + \mathrm{j} \qplain_i$, where $\pplain_i$ and $\qplain_i$ are the net active power and reactive power respectively. The bus voltages are defined in rectangular coordinates, with $\vre_i$ and $\vim_i$ denoting the real and imaginary voltage components, respectively. The voltage magnitudes are given by $v_i=\sqrt{(\vre_i) ^2 + (\vim_i) ^2}$.
	In steady state the electrical network is governed by the nonlinear \ac power flow equations, here given in rectangular form,
		\begin{align}
		\label{eq:PFE_det_full}
		\begin{split}
		    p_i &= \sum_{j \in \mathcal{\nbus}} G_{ij} ( \vre_{i} \vre_{j} + \vim_{i} \vim_{j} ) + B_{ij} ( \vim_{i} \vim_{j} - \vre_{i} \vim_{j} ),\\
		    q_i &= \sum_{j \in \mathcal{\nbus}} G_{ij} ( \vim_{i} \vim_{j} - \vre_{i} \vim_{j} ) - B_{ij} ( \vre_{i} \vre_{j} + \vim_{i} \vim_{j} ),
		\end{split}
		\end{align}
		for all buses $i \in \mathcal{\nbus}$. The matrix $Y = G + \mathrm{j} B \in \mathbb{C}^{\nbus \times \nbus}$ is the bus admittance matrix, which accounts for bus and line shunts as well as transformer tap ratios.
		For ease of presentation, the \ac power flow equations~\eqref{eq:PFE_det_full} are written as a nonlinear system of algebraic equations $g: \mathbb{R}^{4 \nbus} \rightarrow \mathbb{R}^{2 \nbus}$
		\label{eq:PFE_model_det}		
		\begin{equation}
		\label{eq:PFE_det}
		       g(\pplain, \qplain, \vre, \vim) = 0.
		\end{equation}
		In~\eqref{eq:PFE_det}, the $i$th element of $\pplain, \qplain, \vre, \vim \in \mathbb{R}^{\nbus}$ is $\pplain_i, \qplain_i, \vre_i, \vim_i$ for all buses $i \in \mathcal{\nbus}$.
		For simplicity of notation, we assume each bus $i \in \mathcal{\nbus}$ connects to one controllable generation unit $\p_i$ and one uncontrollable power injection $\punc_i$,\footnote{Multiple units at one bus can be easily handled by using matrices that map each generator or uncertainty source to their respective buses.} 
		\begin{align}
		\label{eq:nodal_injections}
		\pplain_i = \p_i + \punc_i, ~		\qplain_i = \q_i + \qunc_i, \quad \forall i \in \mathcal{\nbus}.
		\end{align}

\subsection{Power System Model with Uncertainty}
The power systems model \eqref{eq:PFE_det_full}-\eqref{eq:nodal_injections} assumes a given and fixed set of power injections.
However, power systems operation is influenced by uncertain factors such as fluctuations in temperature, wind speeds, or solar irradiation, which translate into uncertainty in system loading and renewable energy generation.
In this paper, we differentiate between the exogenous drivers of the uncertainty, such as temperature or solar irradiation, and other random quantities that are functions of these exogenous drivers, such as load or solar \textsc{pv} production. The exogenous drivers are modelled through a generic random vector $\omega = [\omega_1, \hdots, \omega_{\nomega}]^\top$ with $\nomega \in \mathbb{N}$, and a corresponding set of possible realizations $\Omega \subset \mathbb{R}^{\nomega}$. This random vector $\omega$ is referred to as the \emph{stochastic germ}.
To account for uncertainty in the \ac power flow equations, the uncontrollable power injections $\punc_i$ and/or $\qunc_i$ at bus $i \in \mathcal{\nbus}$ are modelled as random variables that are (known) functions of the stochastic germ $\omega$
\begin{subequations}
    \label{eq:UncertaintyModel}
	\begin{align}
	    \puncRV_i \triangleq  \puncRV_i(\omega),~
	    \quncRV_i \triangleq  \quncRV_i(\omega), \quad
	    \forall i \in \mathcal{N}.
	\end{align}
	In our notation, sans-serif variables such as $\puncRV_i$, $\quncRV_i$ represent random variables. Realizations of these random variables for a given outcome $\omega \in \Omega$ are written as $\punc_i(\omega)$ or $\qunc_i(\omega)$ to emphasize the functional dependency on the stochastic germ $\omega$, although often $\punc, \qunc$ will be used for compactness of notation.
	The size $\nomega$ of the stochastic germ $\omega$ might be significantly lower than the size $N$ of the uncertainties $\puncRV$, or $\quncRV$. For example, consider temperature as a driver of load uncertainty: If a region is hit by a cold spell or a particularly hot day, the variation in temperature tends to affect many loads in the region, albeit to different degrees.
	
	
	In the following we assume that 
	all occurring random variables have finite variance
    \begin{align}
    \label{eq:FiniteVariance}
	    \var{\omega_j}, \var{\puncRV_i}, \var{\quncRV_i} < \infty, \quad \forall j \in \{ 1,\hdots,\nomega \}, \forall i \in \mathcal{N},
	\end{align}
	\end{subequations}
	where $\var{\cdot}$ denotes the variance.
	This assumption holds in the context of power systems operations, since all quantities are bounded by practical limits such as installed power capacity.
	The assumption~\eqref{eq:FiniteVariance} allows us to handle fairly general cases without imposing any restrictive assumptions on the uncertainty distributions, for example Gaussian.

	

A consequence of the uncertainty model \eqref{eq:UncertaintyModel} is that \emph{all} variables describing the network ($\pplain,\qplain,\vre,\vim$) become random vectors $(\rv{\pplain}(\omega),\rv{\qplain}(\omega), \vreRV(\omega), \vimRV(\omega))$.
	In other words, different realizations $\omega \in \Omega$ define different realizations of the uncertain injections $(\punc_i(\omega),\qunc_i(\omega))$ according to the model~\eqref{eq:UncertaintyModel}.
	This again leads to different realizations of the net active/reactive powers $(\pplain(\omega),\qplain(\omega))$ and voltages $(\vre(\omega),\vim(\omega))$, consistent with the system behavior described by the \ac power flow equations~\eqref{eq:PFE_det}.
	Put differently, the uncertainties $(\puncRV(\omega),\quncRV(\omega))$ are propagated through the power flow equations~\eqref{eq:PFE_det}
	\begin{equation}
	\label{eq:RealizationOfUncertainty}
	    \omega 
	    ~ \overset{\text{\eqref{eq:UncertaintyModel}}}{\Rightarrow} ~
	    \begin{pmatrix}
	        \puncRV(\omega), \quncRV(\omega)
	    \end{pmatrix}
	    ~ \overset{\text{\eqref{eq:PFE_det}}}{\Rightarrow} ~
	    \begin{pmatrix}
	        \rv{\pplain(\omega)},  \rv{\qplain(\omega)},  \vreRV(\omega),  \vimRV(\omega)
	    \end{pmatrix}\!.
	\end{equation}
	Mathematically, the random variables from~\eqref{eq:RealizationOfUncertainty} are defined by the \ac power flow
	\begin{subequations}
		\label{eq:PFE_PerSample}
		\begin{numcases}{\forall \omega \in \Omega:}
		~ 0 = g(\rv{\pplain}(\omega), \rv{\qplain}(\omega), \vreRV(\omega), \vimRV(\omega)),\\
		~ \rv{\pplain}(\omega) = \pRV(\omega) + \puncRV(\omega),\\
		~ \rv{\qplain}(\omega) = \qRV(\omega) + \quncRV(\omega).
		\end{numcases}
	\end{subequations}
    The set of equations~\eqref{eq:PFE_PerSample} can be interpreted as an algebraic equations on random variables. 
    The formulation \eqref{eq:PFE_PerSample} requires the \ac power flow equations to hold for arbitrary uncertainty realizations $\omega$.
    The challenge of modelling the behavior of the nonlinear system and enforcing the \ac power flow equations under uncertainty is a key aspect addressed in this paper.

	\subsection{Chance-constrained Optimal Power Flow}
	\label{sec:CC_OPF}
	In our formulation the goal of chance-constrained \opf  (\ccopf) is to minimize the expected cost of generation, while satisfying the \ac power flow equations for any realization of uncertainty and while guaranteeing that engineering constraints such as voltage magnitude and line current limits will hold up to a pre-specified probability, i.e.
\allowdisplaybreaks
		\begin{subequations} 
		\label{eq:CCOPF}
		\begin{align}
		\underset{\pRV_i, \qRV_i, \vreRV_i, \vimRV_i}{\operatorname{min}}~  & \evbig{ \sum_{i \in \mathcal{\nbus}}  f_i(\pRV_i) } \quad \mathrm{subject\:to}
		\label{eq:ACsOPF_Cost}
		\\
		\label{eq:PFE_RandomVariable}
		&	\hspace{\adjustOPF}
		    g(\rv{p}, \rv{q}, \vreRV, \vimRV) = 0,
		\\
		\label{eq:PFE_pqRV}
		&	\hspace{\adjustOPF}
		    \rv{\pplain} = \pRV + \puncRV, \,
		&&  \hspace{\adjustOPFmod}
		    \rv{\qplain} = \qRV + \quncRV, 
		\\
		\label{eq:CC_Pg}
		& \hspace{\adjustOPF}
		    \mathbb{P}\!\left(  \rv{x} \geq x^{\text{min}} \right) \geq 1 - \varepsilon_{\rv{x}},\, && \hspace{\adjustOPFmod}\rv{x} \in \{\pRV_i, \qRV_i, \vRV_i\},\\
		& \hspace{\adjustOPF}
		\mathbb{P}\!\left(  \rv{x} \leq x^{\text{max}} \right) \geq 1 - \varepsilon_{\rv{x}},\, && \hspace{\adjustOPFmod}\rv{x} \in \{\pRV_i, \qRV_i, \vRV_i\},
		\label{eq:CC_Qg}
		\\
		\label{eq:CC_Ilmag}
		& \hspace{\adjustOPF}
		    \mathbb{P} (  \iRV_{i \text{\unaryminus} j} \leq i_{i \text{\unaryminus} j}^{\text{max}} ) \geq 1 - \varepsilon_{\iplain},
		\\
		\label{eq:VoltageReference}
		& \hspace{\adjustOPF}
		    \vimRV_{i_{\theta V}} = 0,
		&& \hspace{\adjustOPFmod}
		    \forall i \in \mathcal{\nbus}, \: \forall ij \in \mathcal{L}.
		\end{align}
	\end{subequations}	
	Problem~\eqref{eq:CCOPF} minimizes the expected cost of active power generation~\eqref{eq:ACsOPF_Cost}.
	Constraints~\eqref{eq:PFE_RandomVariable}, \eqref{eq:PFE_pqRV} are the power flow equations in terms of random variables, see~\eqref{eq:PFE_PerSample}; in other words, the \ac power flow equalities hold for all realizations of the uncertainties.
	We consider technical limits on the generator active power~$\pRV_i$ and reactive power~$\qRV_i$, as well as constraints on the voltage magnitudes~$\vRV_i$ in \eqref{eq:CC_Pg}, \eqref{eq:CC_Qg}, and line current magnitudes~$\iRV_{i \text{\unaryminus} j}$ in \eqref{eq:CC_Ilmag}.\footnote{{Based on the complex line current from line $i$ to line $j$ which is $i_{i-j} = y_{ij}(e_k - e_m) + \mathrm{j} b_{ij}^{\text{sh}} e_i$ with the complex voltages $e_i, e_j$, branch admittance $y_{ij}$, and shunt susceptance~$b_{ij}^{\text{sh}}$.}}
	These constraints are enforced as \emph{chance constraints} with respective acceptable violation probabilities $\varepsilon_{\pplain}, \varepsilon_{\qplain}, \varepsilon_{\vplain}, \varepsilon_{\iplain} \in [0,1]$.
	The voltage angle reference is set to zero for all realizations of the uncertainty by~\eqref{eq:VoltageReference}.
	An implicit assumption in this paper is that there exists a (high-voltage) solution of the power flow equations for all uncertainty realizations, and that~\eqref{eq:PFE_RandomVariable}, \eqref{eq:PFE_pqRV} models this (high-voltage) solution.
	Further, we observe that the generation dispatch will change as a function of the uncertainty realization, giving rise to an infinite number of variables in formulation~\eqref{eq:CCOPF}. While we discuss how we obtain a finite dimensional representation of the optimization problem, the formulation applied in this paper does not assume any particular form of generator response. Specifically, it is not limited to, e.g., automatic generation control or other affine control policies common in the literature \cite{Vrakopoulou12}.



\subsection{Solution Approach}
In its present form Problem~\eqref{eq:CCOPF} seems intractable owing to infinite-dimensional decision variables, infinite-dimensional equality constraints, and chance constraints that are numerically challenging to evaluate.
We tackle these challenges by expanding all random variables appearing in Problem~\eqref{eq:CCOPF}, using a polynomial basis that is orthogonal with respect to the probability measure~$\mathbb{P}$.
	This approach is called polynomial chaos expansion (\pce) \cite{Sullivan15book,Xiu10book}.
	Polynomial chaos allows to reduce the infinite-dimensional constraints in \eqref{eq:CCOPF} to a set of algebraic equations in the coefficients of the  basis polynomials (so-called Galerkin projection \cite{Sullivan15book, Xiu10book}).
	This way, the infinite-dimensional constraints can be satisfied up to arbitrary numerical accuracy by choosing an appropriately large degree of the polynomial basis, thus bypassing the need for linear approximations, restrictive assumptions on the uncertainty, and/or Monte-Carlo simulations.
	Furthermore, polynomial chaos also allows to compute moments of random variables efficiently without having to sample.
	Thus, it allows to formulate the tasks of uncertainty propagation, moment computation, and optimization elegantly as a single problem;
	i.e. we reformulate~\eqref{eq:CCOPF} as a one-shot finite-dimensional optimization problem.
	


	\section{Introduction to Polynomial Chaos Expansion}
	\label{sec:PCE}
	This section gives a brief overview of polynomial chaos expansion, focusing on its advantages for \ccopf.
	Clearly, the here-given introduction to \pce is non-exhaustive; we refer to \cite{Sullivan15book, Xiu10book} for a more detailed treatment. 
	\subsection{Polynomial Chaos Expansion} \label{subsec:pce}
	Polynomial chaos is a Hilbert space method for random variables that allows a structured representation of uncertainties in terms of deterministic, so-called \pce coefficients.
	Consider $\nomega$ independent random variables $\rv{\omega}_i$ of finite variance for $i = 1,\hdots, \nomega$.
	The random vector $\rv{\omega} \cong [\rv{\omega}_1,\hdots, \rv{\omega}_{N_{\omega}}]^\top$ is called the stochastic germ. 
	Consider the $\nomega$-variate polynomials $\{\basisfun_k \}_{k=0}^{\infty}$ that are orthogonal with respect to the  probability measure $\mathbb{P}(\omega)$, such that
	\begin{equation}
	\label{eq:orthogonality}
	\ev{\basisfun_\ell \basisfun_k} = \langle \basisfun_\ell, \basisfun_k \rangle = \int \basisfun_\ell(\omega) \basisfun_k(\omega)  \mathrm{d}\mathbb{P}(\omega) = \gamma_\ell \delta_{\ell k},
	\end{equation}
	for all $\ell,\, k \in \mathbb{N}_0$.
	In~\eqref{eq:orthogonality} the scalar $\gamma_\ell$ is positive, and $\delta_{\ell k}$ is the Kronecker-delta.
	Notice that every polynomial $\basisfun_k = \basisfun_k(\omega)$ is itself a random variable.
	The orthogonal polynomials are indexed such that their degrees are non-decreasing.
	We define $\basisfun_0 = 1$ as the zero-order polynomial.
	Polynomial chaos expansion allows any real-valued random variable of finite variance that is a function of the stochastic germ $\omega$ to be expressed as a linear combination of the orthogonal polynomials $\{\basisfun_k \}_{k=0}^{\infty}$.
	Specifically, the \pce of the random variable $\hat{\rv{x}}$ is given by
	\begin{subequations}
		\begin{align}
		\label{eq:PCE_infty}
		\hat{\rv{x}} &=  	\sum_{k=0}^\infty x_k \basisfun_k \text{\: with \:} x_k = \frac{\langle \rv{x}, \basisfun_k \rangle}{\langle \basisfun_k, \basisfun_k \rangle} \in \mathbb{R}.
	\end{align}
	The scalars $x_{k}$ are the so-called \pce coefficients.\footnote{For an $\mathbb{R}^n$-valued random vector $\rv{z}$, the \pce representation is obtained by taking the \pce of every component, such that we obtain $\mathbb{R}^n$-valued vectors of \pce coefficients $z_{k} \in \mathbb{R}^n$.}
	\end{subequations}
	For numerical implementations the infinite sum \eqref{eq:PCE_infty} is truncated after $K + 1 \in \mathbb{N}$ terms. In this case, we obtain an approximation $\rv{x}$ of the original random variable $\hat{\rv{x}}$,
	\begin{subequations}
		\label{eq:PCE_complete}	
		\begin{align}
		\hat{\rv{x}} \approx \rv{x} &= \sum_{k \in \mathcal{K}} x_k \basisfun_k \quad \text{with} \quad \mathcal{K} = \{0, \hdots, K\}.
		\end{align}
	\end{subequations}
	The truncation error $\|\hat{\rv{x}} - \rv{x}\| $ is orthogonal to $\rv{x}$ and decays to zero for $K \rightarrow \infty$ in the induced norm $\| \cdot \|$, see \cite{Sullivan15book, Xiu10book}.
	
	\subsection{Uncertainty Propagation and Moment Computation}
	We describe how \pce can be used to (i) propagate uncertainties through (nonlinear) equations, and to (ii) compute moments of output variables. 
    \subsubsection{Uncertainty Propagation}
    \label{sec:GalerkinProjection}
    Consider a given random vector $\rv{x}$ that is mapped/propagated to the random vector $\rv{y}$ according to $h(\rv{x},\rv{y})=0$.
	If the \pce coefficients $x_k$ of $\rv{x}$ are known, the coefficients $y_k$ of $\rv{y}$ can be determined by (intrusive) Galerkin projection; i.e. by projecting onto all orthogonal bases functions $\basisfun_j$ \cite{Sullivan15book, Xiu10book}
	\begin{equation}
	\begin{aligned}
	\label{eq:GalerkinProjection}
	    \forall j \in \mathcal{K}: ~ \, 0 &= \left\langle h\left( \displaystyle\sum_{k \in \mathcal{K}} x_k \basisfun_k, \displaystyle\sum_{k \in \mathcal{K}} y_k \basisfun_k  \right) , \basisfun_j \right\rangle \quad  \\
	    & =: h_j(x_0,\hdots,x_K,y_0,\hdots,y_K).
	\end{aligned}
	\end{equation}
	Hence, the Galerkin projection~\eqref{eq:GalerkinProjection} allows to solve the stochastic problem $h(\rv{x},\rv{y})=0$ by means of $(K{+}1)$ \emph{deterministic} and tractable relations $h_j(\cdot)$.
	The projection error attains a minimum in the induced norm $\| \cdot \|$ and decays to zero for $K \to \infty$ \cite{Sullivan15book, Xiu10book}.
\subsubsection{Computation of Moments}
\label{sec:ComputationOfMoments}
    The moments of a random variable $\rv{x}$ can be expressed as deterministic functions of the \pce coefficients $x_k$.
	For example, the expectation $\ev{\rv{x}}$ and the standard deviation $\sigma[\rv{x}]$ are
	\begin{equation}
	\label{eq:PCE_Moments}
	\ev{\rv{x}} = x_0, ~ \sigma[\rv{x}] = \var{\rv{x}}^{1/2} = \Big(\displaystyle\sum_{k \in \mathcal{K} \setminus \{ 0 \}} \gamma_k x_k^2 \Big)^{1/2}\!\!,
	\end{equation}
	which follows from orthogonality of the basis according to~\eqref{eq:orthogonality}.

\subsection{Construction of Polynomial Basis}
\label{sec:ConstrcutionOfPolynomialBasis}
\begin{table*}
		\caption{Reformulations of power flow equations and moments in terms of \pce coefficients.}
		\vspace{\tablecaptionsep}
		\renewcommand{\arraystretch}{1.2}
		\label{tab:PCE_Reformulations}
		\begin{center}
			\begin{tabular}{l}
				\toprule
				Rectangular power flow in terms of \pce coefficients with $i \in \mathcal{N}$, $k \in \mathcal{K}$\\
				\hline
                $\langle \basisfun_k {,} \basisfun_k\rangle (\p_{i,k}  - \punc_{i,k})=  \sum_{j \in \mathcal{\nbus}} \sum_{k_1, k_2 \in \mathcal{K}} \langle \basisfun_{k_1} \basisfun_{k_2}, \basisfun_{k} \rangle ( G_{ij} (\vre_{i,k_1} \vre_{j,k_2} + \vim_{i,k_1} \vim_{j,k_2}) + B_{ij} (\vim_{i,k_1} \vre_{j,k_2} - \vre_{i,k_1} \vim_{j,k_2}) ) $ \\
				$ \langle \basisfun_k {,} \basisfun_k\rangle (\q_{i,k} - \qunc_{i,k})=  \sum_{j \in \mathcal{\nbus}} \sum_{k_1, k_2 \in \mathcal{K}} \langle \basisfun_{k_1} \basisfun_{k_2}, \basisfun_{k} \rangle ( G_{ij} ( \vim_{i,k_1} \vre_{j,k_2} - \vre_{i,k_1} \vim_{j,k_2}) - B_{ij} ( \vre_{i,k_1} \vre_{j,k_2} + \vim_{i,k_1} \vim_{j,k_2}) ) $ \\
				\hline
				Moments of squared line current magnitudes with $ij \in \mathcal{L}$, $\vre_{ij,k} = \vre_{i,k} - \vre_{j,k}$, $\vim_{ij,k} = \vim_{i,k} - \vim_{j,k}$ \\
				$\ev{\iRV_{i\text{\unaryminus}j}^2} = |y_{ij}^{\text{br}}|^2 \sum_{k \in \mathcal{K}}  \langle \basisfun_k {,} \basisfun_k\rangle ( (\vre_{ij,k} )^2 + ( \vim_{ij,k} )^2 )$  \\
				$ \sigma[ \iRV_{i\text{\unaryminus}j}^2 ]^2 = |y_{ij}^{\text{br}}|^4 \sum_{k_1, k_2, k_3, k_4  \in \mathcal{K}} \langle \basisfun_{k_1} \basisfun_{k_2} \basisfun_{k_3}{,} \basisfun_{k_4}\rangle ( \vre_{i,k_1} \vre_{ij,k_2} \vre_{i,k_3} \vre_{ij,k_4} + 2 \vre_{ij,k_1} \vre_{ij,k_2} \vim_{ij,k_3} \vim_{ij,k_4} + \vim_{ij,k_1} \vim_{ij,k_2} \vim_{ij,k_3} \vim_{ij,k_4})  - \ev{\iRV_{i\unaryminus j}^2}^2$ \\
				\hline
				Moments of squared voltage magnitudes with $i \in \mathcal{\nbus}$ \\
				$\ev{\vRV_{i}^2} = \sum_{k \in \mathcal{K}}  \langle \basisfun_k {,} \basisfun_k\rangle ( (\vre_{i,k})^2 + (\vim_{i,k})^2 )  $ \\
				$ \sigma[ \vRV_{i}^2 ]^2 = \sum_{k_1, k_2, k_3, k_4  \in \mathcal{K}} \langle \basisfun_{k_1} \basisfun_{k_2} \basisfun_{k_3}{,} \basisfun_{k_4}\rangle ( \vre_{i,k_1} \vre_{i,k_2} \vre_{i,k_3} \vre_{i,k_4} + 2 \vre_{i,k_1} \vre_{i,k_2} \vim_{i,k_3} \vim_{i,k_4} + \vim_{i,k_1} \vim_{i,k_2} \vim_{i,k_3} \vim_{i,k_4})  - \ev{\vRV_{i}^2}^2$ \\
				\bottomrule
			\end{tabular}
		\end{center}
		\vspace{\adjustlength}
	\end{table*}
For each component $\omega_i$ of the stochastic germ $\omega$, let  $ \{ \basisfun_k^{(i)} \}_{k=0}^{N_d}$  be the univariate basis of orthogonal polynomials $\basisfun_k^{(i)}(\omega_i)$ with respect to $\mathbb{P}(\omega_i)$, with the degree of $\basisfun_k^{(i)}$ equal to $k$. Then the multivariate \pce basis with respect to the combined stochastic germ $\rv{\omega} \cong [\rv{\omega}_1,\hdots, \rv{\omega}_{N_{\omega}}]^\top$ of maximum degree $N_d$ is given by
\begin{align}
    \label{eq:NumberOfBasisPolynomials}
    \{\basisfun_k\}_{k=0}^{K} \equiv \Big\{ \textstyle\prod_{i=1}^{N_{\omega}} \basisfun_{k_i}^{(i)} \ : 0 \leq \sum_{i}k_i \leq N_d \Big\}.
\end{align}
Notice that the \emph{dimension} of the multivariate basis $\{\basisfun_k\}_{k=0}^{K}$
is given by $ K + 1={(\nomega+N_d)!}/{(\nomega ! N_d!)}$.

%
{To better explain how a multivariate basis is constructed from a set of univariate bases, we provide a simple example.	
	\begin{exmpl}[Bivariate basis of degree at most 2]
	\label{exmpl:BivariateBasis}
	Consider a bivariate stochastic germ $\omega$ with $\nomega = 2$.
	For each $\omega_i$ let the respective univariate basis $ \{ \basisfun_k^{(i)} \}_{k=0}^{K_i} $ have degree $N_d = 2$, where $i \in \{1, 2\}$.
	Each univariate basis has dimension $K_i + 1 = 3$ such that $\{ \basisfun_k^{(1)} \}_{k=0}^{2} = \{ 1, \basisfun_1^{(1)}, \basisfun_2^{(1)} \}$, and $\{ \basisfun_k^{(2)} \}_{k=0}^{2} = \{ 1, \basisfun_1^{(2)}, \basisfun_2^{(2)} \}$, from which the bivariate basis of degree at most $N_d = 2$ is constructed as
	\[
	\{ \underbrace{1}_{\operatorname{deg} = 0}, \underbrace{\basisfun_1^{(1)}, \basisfun_1^{(2)}}_{\operatorname{deg} = 1}, \underbrace{\basisfun_2^{(1)}, \basisfun_1^{(1)}\basisfun_1^{(2)}, \basisfun_2^{(2)}}_{\operatorname{deg}=2} \} =:  \{\basisfun_k \}_{k=0}^K.
	\]
	Hence $K = 5$, which is in accordance with~\eqref{eq:NumberOfBasisPolynomials}.
	\hfill $\square$
	\end{exmpl}
	}
	
	It is desirable to choose a \pce basis that allows an exact representation of a random variable at a low polynomial degree $N_d$, as the dimension $(K{+}1)$---and hence the computational burden---grows rapidly with the degree $N_d$.
	For several univariate continuous random variables the corresponding orthogonal bases are well known and can be used off-the-shelf.
	For example, Hermite polynomials correspond to Gaussian distributions, Jacobi polynomials to Beta distributions, Laguerre polynomials to Gamma distributions \cite{Xiu10book}.
	These random variables admit an \emph{exact} univariate \pce with a basis of maximum degree $N_d=1$ in the respective bases of dimension $K {+} 1 = 2$.
	In other words, random variables that follow Gaussian/Beta/Gamma distributions require only two \pce coefficients to be modelled \emph{exactly}, i.e. there is no error in~\eqref{eq:PCE_complete}.
	In case the random variable is \emph{not} Gaussian/Beta/Gamma, but reasonably similar, the respective bases may still be used at the expense of having to add higher-order coefficients.
	For arbitrary random variables of finite variance it is still possible to find the orthogonal basis such that two \pce coefficients suffice to model the uncertainty exactly.
	
	Mathematically, this amounts to constructing polynomials that are orthogonal with respect to the probability density that describes the uncertainty. This procedure is applicable to both discrete and continuous densities \cite{Sullivan15book}. If the uncertainty is instead described in terms of samples (for example historical data samples), one can proceed by first fitting a density function to the data points. Subsequently, the basis can be constructed either through Gram-Schmidt orthogonalization, the Stieltjes procedure, or the Chebyshev algorithm \cite{Gautschi1982}.

\section{Chance-constrained \opf Using \pce}
\label{sec:CCOPF_UsingPCE}
Having introduced \ccopf in Section~\ref{sec:ProblemFormulation}, and \pce in Section~\ref{sec:PCE}, we now  reformulate the \ccopf Problem~\eqref{eq:CCOPF} as a one-shot optimization problem, namely Problem~\eqref{eq:CCOPF_PCE}.

\subsubsection{Power Injection Uncertainty via PCE}
As seen in Section~\ref{sec:PCE}, a continuous random variable can be represented in terms of its deterministic \pce coefficients. Given a polynomial basis $\{ \basisfun_{k} \}_{k \in \mathcal{K}}$ with $\mathcal{K} = \{ 0, \hdots, K \}$ that is orthogonal with respect to the probability measure $\mathbb{P}(\omega)$, we can represent the nodal power injection uncertainty from \eqref{eq:UncertaintyModel} by
\begin{subequations}
	\begin{equation}
		\puncRV_i = \sum_{k \in \mathcal{K}} \punc_{i,k} \basisfun_{k}, ~
		\quncRV_i = \sum_{k \in \mathcal{K}} \qunc_{i,k} \basisfun_{k} \quad \forall i \in \mathcal{\nbus},
	\end{equation}
\end{subequations}
where $\puncRV, \quncRV$ may follow any distribution with finite variance.

\subsubsection{Uncertainty Propagation for AC Power Flow}  \label{subsec:PCE_UQ}
As described in Section~\ref{sec:PowerSystemModel}, uncertainty in  power injections leads to all network variables $(\pplain,\qplain, \vre, \vim)$ behaving as $\nbus$-valued random vectors $(\rv{\pplain},\rv{\qplain}, \vreRV, \vimRV)$.
We hence model all the network variables using \pce in a common multivariate basis,
\begin{align}
	\label{eq:PCE_ForAllVariables}
	\rv{x} = \sum_{k \in \mathcal{K}} x_k \basisfun_{k} \quad \forall \rv{x} \in \{ \rv{\pplain}_i, \rv{\qplain}_i, \vreRV_i, \vimRV_i, \pRV_i, \qRV_i \}, \: \forall i \in \mathcal{\nbus},
\end{align}
where $x_k$ is the $k$th \pce coefficient of the variable $\rv{x}$, while the basis polynomials $\{ \basisfun_k  \}_{k \in \mathcal{K}}$ are the same for all variables.
Characterizing all the network random variables involves uncertainty propagation through the nonlinear \ac power flow equations, which is in general  challenging.
As described in Section~\ref{sec:GalerkinProjection} it is a main advantage of \pce that it allows to perform this task efficiently using Galerkin projection.
The Galerkin projection~\eqref{eq:GalerkinProjection} applied to the linear equality constraints~\eqref{eq:PFE_pqRV} and \eqref{eq:VoltageReference} is straightforward,
\begin{equation}
    \label{eq:GalerkinProjection_p}
	    \pplain{_k}  = \p_k + \punc_k,~
	    \qplain{_k}  = \q_k + \qunc_k,~\vim_{i_{\theta V},k} = 0,
	    \quad \forall k \in \mathcal{K}.
\end{equation}
Notice that the Galerkin projections \eqref{eq:GalerkinProjection_p}
are \emph{exact}, i.e., the projection errors are zero.
Finally, Galerkin projection is applied to the \ac power flow equations~\eqref{eq:PFE_RandomVariable}, {following the approach in \cite{Muehlpfordt16b}}.
The resulting $2 \nbus (K{+}1)$ deterministic equations are listed in Table~\ref{tab:PCE_Reformulations}.
The Galerkin-projected power flow remains structurally equivalent to deterministic power flow from~\eqref{eq:PFE_det_full}, i.e. the equations are quadratic in the real/imaginary parts of the bus voltages and their sparsity pattern is preserved.
The scalar products $\langle \basisfun_{k_1} \basisfun_{k_2}, \basisfun_{k} \rangle$ from Table~\ref{tab:PCE_Reformulations} can be computed offline using Gauss quadrature.
As described in Section~\ref{subsec:pce}, truncating the \pce at finite $K$ incurs a truncation error.\footnote{
This error is not related to any approximation of the \ac power flow. It stems from the finite truncation of the employed \pce basis.}
The error can be made as small as desired (see description below~\eqref{eq:PCE_complete}) by increasing $K$ at the cost of increased computational burden.
However, as demonstrated in the case studies (Section~\ref{sec:SimulationStudies}), low maximum degrees and hence low \pce dimensions suffice to satisfy the power flow equations to a practical level of accuracy.


\subsubsection{Cost Function}
We consider convex quadratic costs 
\begin{align}
    f_i(\pRV_i) = c_{2,i} (\pRV_i)^2 + c_{1,i} \pRV_i,
\end{align}
with $c_{2,i} > 0$ for every bus $i \in \mathcal{N}$.
The expected cost $\ev{f_i(\pRV_i)}$ per bus from \eqref{eq:ACsOPF_Cost} written in terms of \pce coefficients becomes
\begin{equation}
	\label{eq:Cost_reformulated}
	   \ev{f_i(\pRV_i)} \overset{\phantom{\text{\eqref{eq:PCE_Moments}}}} = c_{2,i} \sum_{k \in \mathcal{K}} \gamma_k \, (\p_{i,k})^2 + c_{1,i} \p_{i,0}  =: \tilde{f}_i(\p_{i,k}),
\end{equation}
with $\gamma_k = \langle \basisfun_{k}, \basisfun_{k}  \rangle$.
Notice that the cost function~$\tilde{f}_i(\p_{i,k})$ remains quadratic, but with respect to the \pce coefficients.

\subsubsection{Chance Constraint Representation}
\label{sec:CC_Reformulations}
We reformulate the chance constraints \eqref{eq:CC_Pg}-\eqref{eq:CC_Ilmag} based on information about their first two moments \cite{Roald15b,Roald18,Calafiore2006}.
For example, the generation constraint in \eqref{eq:CC_Pg} becomes
\begin{equation}
    	\label{eq::CC_Reformulation_First_Two_Moments}
    	p_{\text{g},i}^{\text{min}} \leq 
    	\ev{\pRV_i} \pm \lambda(\varepsilon_{\pplain}) \textstyle{\sqrt{\var{\pRV_i}}}
    	\leq p_{\text{g},i}^{\text{max}},
\end{equation}
where $\lambda(\varepsilon_{\pplain}) > 0$ is chosen based on knowledge about the random variable $\pRV_i$.
For example, in case $\pRV_i$ is Gaussian, the reformulation~\eqref{eq::CC_Reformulation_First_Two_Moments} is exact with $\lambda(\varepsilon_{\pplain}) = \lambda_\Phi(\varepsilon_{\pplain}) := \Phi^{-1}(1 {-} \varepsilon_{\pplain})$, where $\Phi(\cdot)$ is the cumulative distribution function of a standard Gaussian \cite{Bienstock14, Roald13}.
Owing to the nonlinearity of the \ac power flow, the resulting propagated random variables for \ccopf~\eqref{eq:CCOPF} are, however, non-Gaussian in general.
Regardless, the distribution of those variables is often close to a Gaussian in practice.
This is due to a concentration phenomenon similar to the central limit theorem \cite{dasgupta2006, Roald15b}, making $\lambda_\Phi$ a good heuristic that we employ in the following~\cite{Roald18}.
{In case the Gaussian heuristic is unsatisfactory, other choices of $\lambda$ can be used to enforce so-called distributionally robust chance constraints that hold for a family of probability distributions rather than one specific distribution.
As these choices require weaker assumptions (such as symmetry and/or unimodality of the distribution) they become more conservative \cite{Calafiore2006,Roald15b}.
Alternatively, the parameter $\lambda$ can be chosen numerically via cross-validation or through online adaptive methods \cite{oldewurtel2015}.}
The moment-based reformulation~\eqref{eq::CC_Reformulation_First_Two_Moments} is particularly suitable with \pce as moments can be directly obtained from the \pce coefficients, see~\eqref{eq:PCE_Moments}.
Thus, constraint~\eqref{eq::CC_Reformulation_First_Two_Moments} becomes
\begin{equation}
    p_{g,i}^{\text{min}} \leq \p_{i,0} \pm \lambda(\varepsilon_{\pplain})\textstyle\sqrt{\textstyle\sum_{k \in \mathcal{K} \setminus \{ 0 \}} \gamma_k (\p_{i,k})^2 } \leq p_{g,i}^{\text{max}}.
\end{equation}
The reformulation of the other chance constraints for the generator reactive powers~\eqref{eq:CC_Pg}/\eqref{eq:CC_Qg} follows the same procedure.
The chance constraints for voltage magnitudes $\vRV_{i}$ \eqref{eq:CC_Pg}/\eqref{eq:CC_Qg} and line current magnitudes $\iRV_{i\unaryminus j}$ \eqref{eq:CC_Ilmag} are replaced by constraints on their squared magnitudes and the corresponding first and second moment.
The magnitude chance constraints become
\begin{subequations}
	\label{eq:CC_Reformulation_Magnitudes}
	\begin{align}
	(v_{i}^{\text{min}})^2 \leq \ev{\vRV_{i}^2} \pm \lambda(\varepsilon_{\vplain}) \textstyle{\sqrt{\var{\vRV_{i}^2}}}
	\leq (v_{i}^{\text{max}})^2, \\
	\ev{\iRV_{i \text{\unaryminus} j}^2} + \lambda(\varepsilon_{\iplain}) \textstyle{\sqrt{\var{\iRV_{i \text{\unaryminus}j}^2}}} 
	\leq (i_{i \text{\unaryminus}j}^{\text{max}})^2.
	\end{align}
\end{subequations}
The expressions for the moments are given in Table \ref{tab:PCE_Reformulations}.
The reason for using the moment-based reformulation on $\vRV_{i}^2$ and $\iRV_{i\text{\unaryminus}j}^2$ instead of $\vRV_{i}$ and $\iRV_{i\text{\unaryminus}j}$ is that for the former, the moments can be obtained directly as an analytic function of the moments of $\rv{\vre}$ and $\rv{\vim}$ (Table~\ref{tab:PCE_Reformulations}), whereas for the latter, obtaining the moments will require additional equality constraints.
	\subsubsection{Tractable Reformulation}
	The reformulations~\eqref{eq:PCE_ForAllVariables}-\eqref{eq:CC_Reformulation_Magnitudes} allow to cast the chance-constrained \opf~\eqref{eq:CCOPF} as a finite-dimensional nonlinear program (\textsc{nlp}) with the \pce coefficients as decision variables
\allowdisplaybreaks	
	\begin{subequations} 
		\label{eq:CCOPF_PCE}
		\begin{align}
		\label{eq:CCOPF_PCE_Cost}
		\underset{\p_{i,k}, \q_{i,k}, \vre_{i,k}, \vim_{i,k}}{\operatorname{min}} & ~\sum_{i \in \mathcal{\nbus}} \tilde{f}_i(\p_{i,k})  \quad \mathrm{subject\:to}\\
		\label{eq:CCOPF_PCE_PowerFlow}
		&	\hspace{\adjustOPF}
		\text{Galerkin-projected power flow from Table \ref{tab:PCE_Reformulations}},\\
		\label{eq:CCOPF_PCE_CC_Pg}
		& \hspace{\adjustOPF}
		\pplain_{\text{g},i}^{\text{min}} \leq \pplain_{i,0} \pm \lambda(\varepsilon_{\pplain}) \sigma[\pRV_i] \leq \pplain_{\text{g},i}^{\text{max}},
		\\
		& \hspace{\adjustOPF}
		\qplain_{\text{g},i}^{\text{min}} \leq \qplain_{i,0} \pm \lambda(\varepsilon_{\qplain}) \sigma[\qRV_i] \leq \qplain_{\text{g},i}^{\text{max}},
		\\
        &	\hspace{\adjustOPF}
        (v_{i}^{\text{min}})^2 \leq \ev{\vRV_{i}^2} \pm \lambda(\varepsilon_{\vplain}) \sigma[\vRV_{i}^2] \leq (v_{i}^{\text{max}})^2,
        \\
		\label{eq:CCOPF_PCE_Il}
		&	\hspace{\adjustOPF}
		\hspace{1.5cm}\ev{\iRV_{ij}^2} + \lambda(\varepsilon_{\iplain}) \sigma[\iRV_{i \text{\unaryminus} j}^2] 
		\leq (i_{i \text{\unaryminus} j}^{\text{max}})^2,
        \\
		\label{eq:CCOPF_PCE_Slack}
		&	\hspace{\adjustOPF}
		\vim_{i_{\theta V}, k} = 0, \quad \forall k \in \mathcal{K}, 
		\forall i \in \mathcal{\nbus}, \: \forall ij \in \mathcal{L}.
		\end{align}
	\end{subequations}
	The solution to Problem~\eqref{eq:CCOPF_PCE} allows for straightforward a-posteriori uncertainty propagation by means of a simple function evaluation, see~\eqref{eq:RealizationOfUncertainty}). This can be used, e.g. to determine appropriate generator set-points.
	That is, let $\omega$ be the realization of the uncertainty, then with \pce  a-posteriori uncertainty propagation \eqref{eq:RealizationOfUncertainty} becomes
	\begin{equation}
	\label{eq:RealizationOfUncertainty_PCE}
	   \begin{split}
	       \omega \, {\Rightarrow }
	    \sum_{k \in \mathcal{K}}
	    (
	        \punc_k, \qunc_k
	    )
	    \basisfun_k(\omega)
	     \Rightarrow
	    \sum_{k \in \mathcal{K}}
	     (
	        \pplain_k{\!\!}^\star,\!  \qplain_k{\!\!}^\star,\!  \vre_k{}^\star,\!  \vim_k{}^\star
	    ) \basisfun_k (\omega)
	   \end{split}
	\end{equation}
	where the superscript $(\cdot)^\star$ denotes the solution to \eqref{eq:CCOPF_PCE}.
	To evaluate~\eqref{eq:RealizationOfUncertainty_PCE} means to evaluate the basis polynomials $\basisfun_k$ at the realization $\omega$, and then to multiply by the \pce coefficients---which is computationally cheap.

\section{Case Studies}
\label{sec:SimulationStudies}
Next, we demonstrate the practicability and advantages of \pce for stochastic optimal power flow.
In particular we study the numerical accuracy of the \ac power flow equations for varying maximum degrees; the accuracy of the moments; the empirical violation probability of selected inequality constraints both for in-sample and out-of-sample tests; and the shape of the generation policies.

In the case studies all numbers are given in per-unit (p.u.) for a base \textsc{mva} of 100.
Simulations were carried out on a standard desktop computer with 16\,\textsc{gb\,ram} and an i7-4770 \textsc{cpu}, and implemented in Julia using \textsc{j}u\textsc{mp} with  Ipopt as \textsc{nlp} solver.
{
We initialized the \textsc{nlp}~\eqref{eq:CCOPF_PCE} as follows.
The zero-order \pce coefficients were set equal to the solution of the deterministic \opf problem with the uncertainties set to their expected values, while the higher-order coefficients were set to zero.
The solution of the deterministic~\opf also provided an initial guess for the active set of~\eqref{eq:CCOPF_PCE}.
To improve computational tractability, we applied a constraint generation method to solve the \textsc{nlp}.
First, we solve the \textsc{nlp} including only the constraints that were active for the deterministic problem. Second, we check a posteriori whether the obtained solution satisfies \emph{all} constraints.
If a constraint is violated, we add the violated constraint to the problem, and solve the \textsc{nlp} again.
}


\subsection{Satisfaction of AC Power Flow Constraints} 
\label{sec:SatisfactionOfACPowerFlowConstraints}
As described in Section~\ref{subsec:PCE_UQ},  there is a trade-off between the accuracy of \ac power flow satisfaction and the \pce dimension $(K{+}1)$, which dictates the computational complexity.
Since keeping the order $K$ low is desirable, we investigate what is the smallest \pce degree that is sufficient for the \ac power flow constraints to be satisfied up to a practical level of accuracy.
These tests are performed on the 5-bus and 30-bus test case.


\subsubsection{5-Bus Test Case}
\label{sec:5BusTestCase}
	\begin{figure}
		\centering
		\subfloat[Modified 5-bus test case.\label{fig:5BusTestCase}]{\includegraphics[width=5cm]{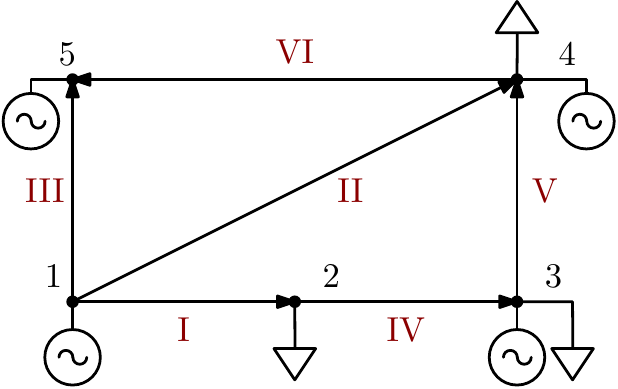}}%
		\quad
		\subfloat[Probability density of demand at bus 2.\label{fig:5Bus_PDF}]{\includegraphics[width=3cm]{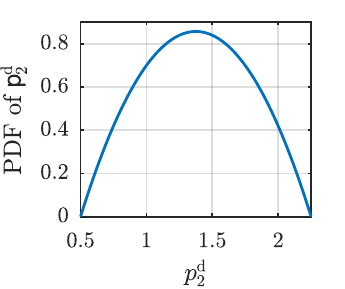}}%
		\caption{5-bus test case and uncertainty model.}
	\end{figure}
\begin{table}
	\centering
	\caption{Parameters for 5-bus test case.}
	\vspace{\tablecaptionsep}
	\renewcommand{\arraystretch}{\rowstretch}
	\label{tab:5Bus_CostCoefficients}
	\begin{tabular}{ccccccc|cc}
		\toprule
		Bus $i$ & $c_{2,i}$ & $c_{1,i}$ & $p_{g,i}^{\text{max}}$ & $q_{g,i}^{\text{max}}$ & $\vplain_{i}^{\text{min}}$  & $\vplain_{i}^{\text{max}}$   & Line $i$-$j$ & $i_{i\text{\unaryminus}j}^{\normalfont \text{max}} $\\
		\midrule
		1 & 14 & 2 & 1.5 & 1.275	& 0.9 & 1.1 &	1-2 & 1.04\\
		2 & -	& -	&- 	& -	& 	0.9 	& 1.1 & 1-5 & 0.87\\
		3 & 11 & 3 & 4.3 & 3.9  	&	0.9 & 1.1 & 2-3 & 0.78\\
		4 & 14 & 4 & 9.9  & 1.5 		& 0.9& 1.1 &	\\
		5 & 13 & 1 & 1.5 & 4.5 		&  0.9 & 1.1 & 	\\
		\bottomrule
	\end{tabular}
	\vspace{\adjustlength}
\end{table}	
We consider a modified version of the 5-bus test from \cite{Zimmerman11}, shown in Figure~\ref{fig:5BusTestCase}. We neglect the shunt elements, consider a quadratic cost function, and assume that the voltage magnitude at the slack bus~4 is constant at one. The line current limits are set equal to the per-unit \textsc{mva} ratings.
Other relevant parameters are summarized in Table~\ref{tab:5Bus_CostCoefficients}.

The active power demand at bus 2 is uncertain and follows a Beta distribution with support $[l, u] = [0.50, 2.25]$ and shape parameters $(\alpha, \beta) = (2, 2)$,
$
\puncRV_2 = - \pdRV_2 ~ \text{with} ~ \pdRV_2 \sim \mathsf{B}( [0.50, 2.25], 2, 2),
$
with $\ev{\pdRV_2} = 1.375$ and $\sigma[\pdRV_2] = 0.391$.
The \pce coefficients are $[\punc_{2,0}, \punc_{2,1}] = 1/(\alpha+\beta)\,[\beta u+\alpha l, u-l]$.
The probability density function is plotted in Figure~\ref{fig:5Bus_PDF}.
Since the uncertainty is one-dimensional, the stochastic germ is equal to $\omega {\sim} \mathsf{B}( [0, 1], 2, 2 )$, and the Jacobi polynomials provide the corresponding orthogonal basis \cite{Xiu10book}.
For the chance constraints, we enforce $\varepsilon = \varepsilon_{\pplain} = \varepsilon_{\qplain} = \varepsilon_{\vplain} = \varepsilon_{\iplain} = 0.1$, and we set the inequality constraint parameter to $\lambda(\varepsilon) = \lambda_\Phi(\varepsilon) = 1.2816$, see Section~\ref{sec:CC_Reformulations}.
\begin{figure}
	\centering
    \includegraphics{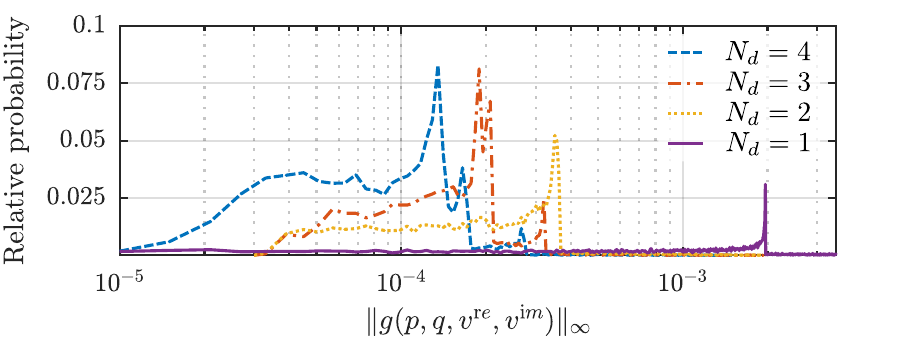}
    \vspace{-7mm}
	\caption{Relative probability of maximum \ac power flow violation for polynomial bases of degree at most $N_d \in \{1, 2, 3, 4\}$ for 5-bus system.}
	\label{fig:5Bus_PFEviolations}
	\vspace{\adjustlength}
\end{figure}
We consider a single univariate source of uncertainty. Hence, the dimension and the maximum degree of the basis are linked by $K {+} 1 = N_d {+} 1$, implying that the number of constraints grows linearly with the maximum degree $N_d$.
We draw 10\,000 realizations of the uncertainty, and for each realization we compute the values of all remaining variables based on~\eqref{eq:RealizationOfUncertainty_PCE}.
To quantify the error in the \ac power flow satisfaction, we compute the $\infty$-norm of~\eqref{eq:PFE_det} for these realizations, and compare it against its ideal value of zero. The relative probability of $\|g(\cdot)\|_{\infty}$ is shown in Figure~\ref{fig:5Bus_PFEviolations}, where the different lines correspond to different maximum degrees $N_d$ of the polynomial basis.
As expected, a larger maximum degree $N_d$---hence a larger \pce dimension---leads to lower maximum \ac power flow violations.
There is a sharp decrease in the power flow inaccuracies from $N_d=1$, with errors between $\text{1\,\textsc{e}-3}$ and $\text{2\,\textsc{e}-3}$, to $N_d=2$  with errors approximately at 4\,\textsc{e}-4.
Further increases in the degree decrease the error even more, but not considerably.
The order of \pce necessary for sufficient accuracy depends on the effective nonlinearity in the power flow equations. For example, in case of \dc power flow, it is known that \pce with degree $N_d = 1$ is exact \cite{Muehlpfordt18b}.
The fact that in our experiments a \pce basis of degree 2 has a small error shows that a degree of 2 is enough to capture the level of nonlinearity of the \ac power flow equations for typical levels of uncertainty.
The solution times for the \textsc{nlp} \eqref{eq:CCOPF_PCE} are 0.45\,s, 0.84\,s, 0.67\,s, 2.6\,s for maximum degrees 1, 2, 3, 4, respectively.

\subsubsection{30-bus Test Case}
\label{sec:30BusTestCase}
We consider a modified version of the 30-bus test case \cite{Zimmerman11}. The shunt elements are neglected for simplicity, and the voltage magnitude at slack bus~1 is assumed constant at one. 
The line current limits are set to the nominal values of the per-unit line ratings, except for two lines where the capacity is reduced from $16$ to,~$i_{\text{15-23}}^{\normalfont \text{max}} = 11$, and $i_{\text{25-27}}^{\normalfont \text{max}} = 12$.
Reducing the capacity on those two lines makes for a more interesting case, as several line current limits become binding.

We introduce a stochastic germ $\omega$ comprised of four distinct sources of uncertainty, two Beta distributions (one symmetric, one non-symmetric) and two normal distributions as described in  Table~\ref{tab:30_bus_StochasticGerm}.
The stochastic germ $\omega$ is used to represent load uncertainty at six buses $i \in \mathcal{U} = \{ 2, 3, 4, 24, 10, 21 \}$, as listed in the last column of Table~\ref{tab:30_bus_StochasticGerm}.
For each bus $i \in \mathcal{U}$, the uncertain load is modelled as 
 $\puncRV_{i} = - \pdRV_{i} ~~ \text{with} ~~ \ev{\pdRV_{i}} = \pdnom_{i}, ~ \sigma[\pdRV_{i}] = \rsd  \cdot \pdnom_{i},
 $
where $\pdnom_{i}$ is the nominal value of the active power demand taken from the case file \cite{Zimmerman11}, and the relative standard deviation $\rsd > 0 $ describes the standard deviation as a fraction of the nominal load.

First, we fix the relative standard deviation $\rsd = 0.15$ and the risk level $\varepsilon = 0.15$. We verify the satisfaction of the power flow equations for varying maximum degrees $N_d \in \{1, 2, 3\}$, according to the procedure from the 5-bus test case.
Due to space constraints we provide only the maximum power flow violation across all samples in Table~\ref{tab:30Bus_PFEviolations}. Graphically this corresponds to the right-most value in Figure~\ref{fig:5Bus_PFEviolations} of every plotted line.
Table~\ref{tab:30Bus_PFEviolations} supports the findings from the 5-bus system---increasing the degree leads to higher accuracy of the power flow equations and degree $2$ provides sufficient accuracy in practice.
Notice that overall the power flow equations are more accurate for the 30-bus system compared to the 5-bus system.
As the 5-bus system is more meshed, the effect of the uncertainty is greater there---even though there is just a single source of uncertainty.

\begin{table}
	\centering
	\caption{Max. AC power flow violation for 30-bus system.}
	\vspace{\tablecaptionsep}
	\renewcommand{\arraystretch}{\rowstretch}
	\label{tab:30Bus_PFEviolations}
	\begin{tabular}{lrrr}
		\toprule
		Maximum degree $N_d$ & 1 & 2 & 3\\
		Maximum power flow violation & 3.8548\textsc{e}-5 & 3.66973\textsc{e}-6 & 2.34221\textsc{e}-8\\
		\textsc{nlp} time in seconds& 0.6 & 10.6 & 239.2 \\
		\bottomrule
	\end{tabular}
	\vspace{\adjustlength}
\end{table}	

\begin{table*}
	\centering
	\caption{Error in the computed moments for the \pce method (\pce) and the linearization method (lin) for 30-bus system.}
	\label{tab:Moments_PCE}
	\vspace{\tablecaptionsep}
	\renewcommand{\arraystretch}{\rowstretch}
	\begin{tabular}{ccrrrrrrrr}
		\toprule
		\rowcolor{white!100}                    &  &                                 \multicolumn{2}{c}{$\p$}                                 &                                 \multicolumn{2}{c}{$\q$}                                 &                              \multicolumn{2}{c}{$\vplain$}                               &                        \multicolumn{2}{c}{$\iplain_{i\text{\unaryminus}j}$}                          \\
		\multirow{-2}{*}{\ac vs.} & \multirow{-2}{*}{\cellcolor{white}$\rsd$} &               $\| \Delta \mu \|_{\infty}$ &               $\| \Delta \sigma \|_{\infty}$ &               $\| \Delta \mu \|_{\infty}$ &               $\| \Delta \sigma \|_{\infty}$ &               $\| \Delta \mu \|_{\infty}$ &               $\| \Delta \sigma \|_{\infty}$ &               $\| \Delta \mu \|_{\infty}$ &               $\| \Delta \sigma \|_{\infty}$  \\ \midrule
		\cellcolor{gray!25} & 0.05                        &                          1.8\,\textsc{e}-5 &                            0.6\,\textsc{e}-5 &                         1.7\,\textsc{e}-5 &                            0.7\,\textsc{e}-5 &                         0.3\,\textsc{e}-5 &                           0.4\,\textsc{e}-5 &                        5.1\,\textsc{e}-5 &                           3.9\,\textsc{e}-5               \\
		\cellcolor{gray!25}& \cellcolor{white}0.10                        &                        10.1\,\textsc{e}-5 &                            0.4\,\textsc{e}-5 &                         2.0\,\textsc{e}-5 &                            2.0\,\textsc{e}-5 &                         2.2\,\textsc{e}-5 &                            1.0\,\textsc{e}-5 &                        33.4\,\textsc{e}-5 &                            5.7\,\textsc{e}-5\\
		\cellcolor{gray!25}\multirow{-3}{*}{\pce}  & \cellcolor{white}0.15                        &                         2.9\,\textsc{e}-5 &                           19.8\,\textsc{e}-5 &                        10.7\,\textsc{e}-5 &                            6.4\,\textsc{e}-5 &                         3.8\,\textsc{e}-5 &                            1.1\,\textsc{e}-5 &                        19.3\,\textsc{e}-5 &                           12.1\,\textsc{e}-5             \\ \midrule
		\cellcolor{gray!25} &		\cellcolor{white}0.05   &                       431.2\,\textsc{e}-5 &                            4.0\,\textsc{e}-5 &                                     0.131 &                           81.1\,\textsc{e}-5 &                       108.4\,\textsc{e}-5 &                            2.3\,\textsc{e}-5 &                      4901.5\,\textsc{e}-5 &                          104.0\,\textsc{e}-5              \\
		\cellcolor{gray!25} & \cellcolor{white}0.10   &                       411.0\,\textsc{e}-5 &                           12.9\,\textsc{e}-5 &                                     0.136 &                          294.3\,\textsc{e}-5 &                       105.8\,\textsc{e}-5 &                            7.5\,\textsc{e}-5 &                      4812.1\,\textsc{e}-5 &                          196.3\,\textsc{e}-5          \\
		\cellcolor{gray!25}\multirow{-3}{*}{lin. \ac}  & \cellcolor{white}0.15   &                       387.6\,\textsc{e}-5 &                            7.0\,\textsc{e}-5 &                                     0.146 &                          700.3\,\textsc{e}-5 &                       101.8\,\textsc{e}-5 &                           17.1\,\textsc{e}-5 &                      4715.7\,\textsc{e}-5 &                          348.1\,\textsc{e}-5               \\
		\midrule 
		&                           & $\| \ev{\pRV}_{\text{\ac}}   \|_{\infty}$ & $\| \sigma[\pRV]_{\text{\ac}}   \|_{\infty}$ & $\| \ev{\qRV}_{\text{\ac}}   \|_{\infty}$ & $\| \sigma[\qRV]_{\text{\ac}}   \|_{\infty}$ & $\| \ev{\vRV}_{\text{\ac}}   \|_{\infty}$ & $\| \sigma[\vRV]_{\text{\ac}}   \|_{\infty}$ & $\| \ev{\iRV_{i\text{\unaryminus}j}}_{\text{\ac}}   \|_{\infty}$ & $\| \sigma[\iRV_{i\text{\unaryminus}j}]_{\text{\ac}}   \|_{\infty}$   \\
		& \multirow{-2}{*}{Reference}               &                                    0.5800 &                                       0.1132 &                                    0.3829 &                                       0.0038 &                                    1.0792 &                                       0.0012 &                                    0.3951 &                                       0.0100   \\
		\bottomrule
	\end{tabular}
\end{table*}
\begin{table*}
\scriptsize
\centering
\caption{Empirical constraint satisfaction, cost, and expected power flow violation for max. degree $N_d {\in} \{1, 2 \}$ for 30-bus system.}
\label{tab:30Bus_ConstraintSatisfaction}
\vspace{\tablecaptionsep}
\renewcommand{\arraystretch}{\rowstretch}
\begin{tabular}{cc|ccccccc|ccccccc}
\toprule
    & & \multicolumn{7}{c}{Maximum degree $N_d = 1$ } & \multicolumn{7}{c}{Maximum degree $N_d = 2$} \\
    & & & & & & &   & \textsc{pf} violation & & & & & &  & \textsc{pf} violation\\
    \multirow{-2}{*}{$\rsd$}   &   \multirow{-2}{*}{$\varepsilon$} 
    & \multirow{-2}{*}{$p_{\text{g},3}^{\text{max}}$}  & \multirow{-2}{*}{$p_{\text{g},4}^{\text{max}}$}  & \multirow{-2}{*}{$i_{\text{21-22}}^{\text{max}}$} & \multirow{-2}{*}{$i_{\text{15-23}}^{\text{max}}$} & \multirow{-2}{*}{$i_{\text{25-27}}^{\text{max}}$} &  \multirow{-2}{*}{Cost} & $\ev{\cdot}{}$/1\,\textsc{e}-3
    & \multirow{-2}{*}{$p_{\text{g},3}^{\text{max}}$}  & \multirow{-2}{*}{$p_{\text{g},4}^{\text{max}}$}  & \multirow{-2}{*}{$i_{\text{21-22}}^{\text{max}}$} & \multirow{-2}{*}{$i_{\text{15-23}}^{\text{max}}$} & \multirow{-2}{*}{$i_{\text{25-27}}^{\text{max}}$} &  \multirow{-2}{*}{Cost} & $\ev{\cdot}{}$/1\,\textsc{e}-3 \\
    \midrule
     &          0.05  & 0.9495 & 0.9499 & 0.9436 & 0.9514 & 0.9490 & 599.25 & 0.0640 
                      & 0.9494 & 0.9499 & 0.9424 & 0.9526 & 0.9483 & 599.25 & 0.0009  \\
     &          0.10  & 0.9026 & 0.9015 & 0.8954 & 0.8959 & 0.8980 & 599.24 & 0.0618  
                      & 0.9022 & 0.9012 & 0.8953 & 0.8961 & 0.8980 & 599.24 & 0.0007  \\
     \multirow{-3}{*}{0.10} & 0.15 
                      & 0.8514 & 0.8507 & 0.8812 & 0.8515 & 0.8489 & 599.24 & 0.0601
                      & 0.8515 & 0.8506 & 0.8808 & 0.8516 & 0.8487 & 599.24 & 0.0006 \\
     
 \midrule
    &          0.05  & 0.9494 & 0.9499 & 0.9388 & 0.9486 & 0.9475 & 599.38 & 0.1502
                     & 0.9494 & 0.9500 & 0.9381 & 0.9511 & 0.9473 & 599.38 & 0.0042 \\
     &          0.10 & 0.9028 & 0.9015 & 0.8938 & 0.8937 & 0.8969 & 599.36 & 0.1421
                     & 0.9030 & 0.9015 & 0.8927 & 0.8941 & 0.8969 & 599.36 & 0.0029 \\
     \multirow{-3}{*}{0.15} & 0.15 
                     & 0.8514 & 0.8502 & 0.8488 & 0.8427 & 0.8486 & 599.35 & 0.1373
                     & 0.8515 & 0.8501 & 0.8484 & 0.8426 & 0.8485 & 599.35 & 0.0024 \\
\bottomrule
\end{tabular}
\vspace{\adjustlength}
\end{table*}
\subsection{In-sample Tests}
In this section we investigate the capability of \pce to reduce the constraint violation probability to below an acceptable level.
Since our chance constraint reformulation is based on the first and second moments of the uncertainty, we first assess their accuracy when computed with \pce.
We then assess the ability of the method to limit the level of constraint violations.


\subsubsection{Accuracy of Moment Computation}
\label{sec:AccuracyOfMoments}
The quality of the moments is essential for the reformulations of the chance constraints, which is evident from \eqref{eq::CC_Reformulation_First_Two_Moments} and \eqref{eq:CC_Reformulation_Magnitudes}.
We compare the accuracy of the \pce-based moments with moments obtained from the full nonlinear \ac power flow equations, and a linearized version of the \ac power flow equations used in the literature \cite{Roald18}. 
The variables in the \ac power flow equations are divided into \emph{independent} variables---real and imaginary voltage at the slack bus, active power and voltage magnitude at the \textsc{pv} buses and real and reactive power injection at the \textsc{pq} buses---and \emph{dependent} variables, which consist of the rest.
Each realization of the uncertainty $\omega$ fully specifies the \emph{independent} variables: the real and reactive power consumption of the uncertain loads are determined by $\omega$, and the active power and voltage magnitude at the \textsc{pv} buses are adjusted according to the control policy given by \eqref{eq:RealizationOfUncertainty_PCE}.

For the 30-bus test case and a maximum degree of $N_d = 2$, we compare three sets of moments:
(i) The \pce moments $(\ev{\cdot}_{\text{\pce}}, \sigma[\cdot]_{\text{\pce}})$ obtained directly from \eqref{eq:PCE_Moments}.
(ii) The moments for full \ac power flow $(\ev{\cdot}_{\text{\ac}}, \sigma[\cdot]_{\text{\ac}})$ obtained by drawing uncertainty samples, determining the value of the \emph{independent} variables in a similar fashion as for \pce, and then solving the full \ac equations to determine the value of the \emph{dependent} variables. (iii) The moments of the linearized \ac power flow $(\ev{\cdot}_{\text{lin}}, \sigma[\cdot]_{\text{lin}})$ obtained via sampling, but the \emph{dependent} variables are determined using a first-order Taylor approximation around the operating point corresponding to~$\ev{\omega}$.

If sufficiently many samples are used in the Monte Carlo simulations, the moments obtained from the full \ac equations $(\ev{\cdot}_{\text{\ac}}, \sigma[\cdot]_{\text{\ac}})$ can be considered as ground truth.
The quality of the \pce and linearization is given by the $\infty$-norm of the error relative to the \ac solution.
For the active power we compute the error by comparing the expected value via \pce to the expected value  via the full \ac power flow  as
	\begin{align}	\label{eq:Moments_PCE}
	\| \Delta \mu \|_{\infty} &= \| \ev{\pRV}_{\text{\ac}} - \ev{\pRV}_{\text{\pce}}  \|_{\infty}. 
	\end{align}
	The error in the standard deviation $\sigma$ for the \pce method, as well as the expected value and standard deviation for the linearization method are evaluated analogously.

\begin{table}
    \vspace{2mm}
	\centering
	\caption{Stochastic germ and affected buses.}
	\label{tab:30_bus_StochasticGerm}
	\vspace{\tablecaptionsep}
	\renewcommand{\arraystretch}{\rowstretch}
	\begin{tabular}{crrr}
		\toprule
		$\omega_j$ &          Distribution          & Polynomial basis & Affected buses \\ \midrule
		1      & $\mathsf{B}([$0, 1$]$, 2, 2$)$ &      Jacobi      &      2, 3      \\
		2      & $\mathsf{B}([$0, 1$]$, 2, 5$)$ &      Jacobi      &       4        \\
		3      &      $\mathsf{N}($0, 1$)$      &     Hermite      &       24       \\
		4      &      $\mathsf{N}($0, 1$)$      &     Hermite      &     10, 21     \\ \bottomrule
	\end{tabular}
	\vspace{\adjustlength}
\end{table}

Table~\ref{tab:Moments_PCE} summarizes the results for varying relative standard deviations $\rsd$ of the load, and for a total of 10\,000 Monte Carlo samples of the full \ac power flow.
We observe that \pce performs significantly better than the linearized power flow, with \pce giving errors that are orders of magnitude smaller. 
We observe that \pce errors ($\approx$1\,\textsc{e}-5) are also small relative to the reference values in the last row of Table~\ref{tab:Moments_PCE}, implying that \pce is quite accurate for all considered values of the relative standard deviation $\rsd$. 
Table~\ref{tab:Moments_PCE} also shows that the reactive power behaves more nonlinearly than the active power, leading to larger errors in the reactive power estimation for the linearized power flow. Particularly the mean values are poorly estimated. 
Since the accuracy of the reformulated chance constraints requires accuracy in both mean and standard deviation, \pce is expected to be superior in enforcing chance constraints compared to the linearized \ac power flow.

\subsubsection{Chance Constraint Satisfaction} 
Next, we investigate how the maximum degree $N_d$ affects constraint satisfaction.
Table \ref{tab:30Bus_ConstraintSatisfaction} summarizes the results for maximum degrees $N_d \in \{1, 2 \}$, considering relative standard deviations $\rsd$\,$ \in \{ 0.10, 0.15\}$ and violation probabilities $\varepsilon$\,${\in} \{ 0.05, 0.10, 0.15 \}$.
The empirical constraint satisfaction is computed from 10\,000 samples, evaluated using the full \ac power flow equations.
From Table~\ref{tab:30Bus_ConstraintSatisfaction} we observe that for both $N_d = 1$ and $N_d = 2$ there are no significant violations of the chance constraints, i.e., the empirical constraint satisfaction is close to the specified level $1 {-} \varepsilon$. 
There are several smaller inaccuracies i the enforcement of the chance constraints that may be attributed to the uncertainty being non-Gaussian, however, this effect appears to be small. 
The empirical constraint satisfaction for the two maximum degrees $N_d = 1$ and $N_d = 2$ is similar, although $N_d = 1$ yields slightly lower constraint satisfaction compared to $N_d = 2$.
This may be because for $N_d = 1$ we are not able to capture the skewness of the distributions as well as for a maximum degree $N_d =2$.
Table~\ref{tab:30Bus_ConstraintSatisfaction} also shows the expected power flow (\textsc{pf}) violation. Consistent with our results in Section~\ref{sec:SatisfactionOfACPowerFlowConstraints}, the expected power flow violation for the maximum degree $N_d = 1$ is considerably larger compared to $N_d = 2$, e.g. 0.1502\,\textsc{e}-3 vs. 0.0042\,\textsc{e}-3 for the relative standard deviation $\rsd = 0.15$ and the violation probability $\varepsilon = 0.05$.
To summarize, a higher maximum degree $N_d$ ensures more accurate satisfaction of the power flow equations \emph{and} of the chance constraints.

\subsection{Out-of-sample Tests}
\label{sec:OutofsampleTests}
{
Since it is frequently hard to obtain accurate estimates of the probability distributions in practical applications, we are also interested in understanding the out-of-sample performance of the method. To assess the out-of-sample performance, we test the solution of~\eqref{eq:CCOPF_PCE} against distributions that are different from what we assumed when solving~\eqref{eq:CCOPF_PCE}. In the following, we will refer to the uncertainty assumed within~\eqref{eq:CCOPF_PCE} as \emph{modelled uncertainty}, and the uncertainty we test against as \emph{actual uncertainty}.}

{
We compute the solution of~\eqref{eq:CCOPF_PCE} for the stochastic germ from Table~\ref{tab:30_bus_StochasticGerm} with a relative standard deviation~$ \rsd = 0.15$ and a risk level~$\varepsilon = 0.15$, and for varying maximum degrees $N_d \in \{1, 2, 3 \}$.
Each of these three solutions---which have the same modelled uncertainty but vary in the maximum polynomial degree---is stored and tested against two different actual uncertainties with respect to \ac power flow and chance constraint satisfaction.
}
\subsubsection{Correct Distribution -- Inaccurate Standard Deviations}
{
First, we perform out-of-sample test where the actual uncertainty belong to the same family of distributions as the modelled uncertainty (see Table~\ref{tab:30_bus_StochasticGerm}), but where we have an inaccurate estimate of the standard deviations.
In this case, the actual uncertainties still follow Beta and Normal distributions, but we scale the relative standard deviation of the actual uncertainty, such that we have three standard deviations $\{ 0.5s, s, 1.5s\}$, which are smaller, the same and larger than the assumed standard deviation. 
We include the case for which the actual uncertainty is equivalent to the modelled uncertainty to have a means of comparison.
}

{
The left plot of Figure~\ref{fig:30Bus_outofsample_violation} shows the maximum power flow violation among 10\,000 samples for varying maximum degrees $N_d \in \{1, 2, 3\}$ as a function of the relative standard deviation~$\rsd$.
For the same value of the relative standard deviation~$\rsd$ the \ac power flow satisfaction reduces greatly as the maximum degree increases. This is consistent with the results from Section~\ref{sec:SatisfactionOfACPowerFlowConstraints}.
On the other hand, the accuracy of the power flow equations appears fairly insensitive to inaccuracies in the standard deviation~$\rsd$.
For example, for the maximum degree $N_d = 2$ the maximum power flow error is in the range of 1\,\textsc{e}-6 for all values of the relative standard deviation~$\rsd$.}

{
The variation of the empirical chance constraint satisfaction for different standard deviations is shown in the right plot of Figure~\ref{fig:30Bus_outofsample_violation} for the upper branch flow limit of the line connecting buses 21 and 22.
We observe that empirical constraint satisfaction decreases with increasing values of~$\rsd$. 
Conversely, the constraint satisfaction appears is not dependent on the maximum degree~$N_d$. 
It is not surprising that the empirical violation probability is sensitive to the inaccuracies in standard deviations, as the standard deviation enters directly into the expressions for the reformulated chance constraints. The increase in constraint satisfaction to more than $95\,\%$ for $0.5 \rsd$ stems from the smaller support of the actual uncertainty relative to the modelled one.\footnote{Technically, the support is only smaller for the Beta distribution while it remains the real axis for Gaussians regardless of $\nu$, but numerically the samples for the Gaussian distributions will have a smaller support.}
On the other hand, positive perturbations such as $1.5 \rsd$ increase the support beyond what was assumed in the reformulation, hence constraint violations are more frequent.
}

\subsubsection{Inaccurate Distributions -- Correct Standard Deviation}
{
In our second set of experiments, the actual uncertainty matches the first two moments of the modelled uncertainty, but the underlying family of distributions is different.
Specifically, the stochastic germ for the actual uncertainties is set to the uniform distribution $\omega_j \sim \mathsf{U}([0, 1])$ for $j \in \{1, 2, 3, 4\}$ in the first experiment, and to the Beta distribution $\omega_j \sim \mathsf{B}([0,1], \alpha, \beta)$ with shape parameters $\alpha = \beta = 3$ for $j \in \{1, 2, 3, 4\}$ in the second experiment.
Both of these cases are compared to the case for which the actual uncertainty matches the modelled uncertainty.
}

\begin{figure}
	\centering
	\includegraphics{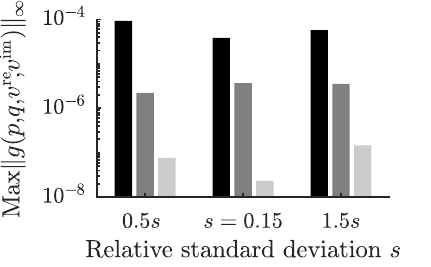}
	\hfill
    \includegraphics{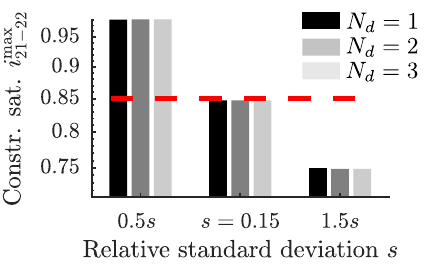}
	\caption{Correct distribution, inaccurate standard deviations -- power flow violation and constraint satisfaction for varying maximum degrees.}
	\label{fig:30Bus_outofsample_violation}
	\vspace{-2.5mm}
\end{figure}

\begin{figure}
	\centering
    \includegraphics{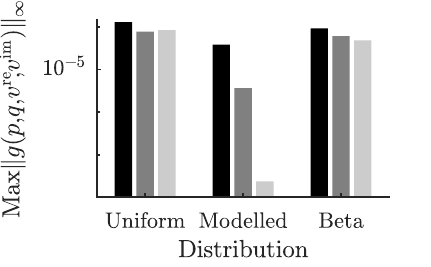}
    \hfill
    \includegraphics{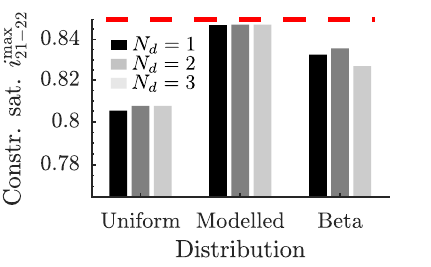}
	\caption{Inaccurate distributions, correct standard deviation -- power flow violation and constraint satisfaction for varying maximum degrees.}
	\label{fig:30Bus_outofsample_violation_distr_robust}
	\vspace{\adjustlength}
\end{figure}

{
The left plot of Figure~\ref{fig:30Bus_outofsample_violation_distr_robust} shows the maximum power flow violation for varying maximum degrees $N_d \in \{1, 2, 3\}$ for the three different actual uncertainties.
The exactness of the \ac power flow still decreases with increasing the maximum degree, but the decrease is not as significant as the decrease seen in Figure~\ref{fig:30Bus_outofsample_violation}.
The empirical constraint satisfaction, however, still seems fairly insensitive to the maximum degree, although the desired level of $1 - \varepsilon = 0.85\, \%$ is met only for the case in which actual and modelled uncertainty coincide.
}

{
Based on the in-sample and out-of-sample tests we draw the following conclusions: (i) Increasing the maximum degree~$N_d$ leads to more accurate \ac power flow, with $N_d = 2$ giving a good trade-off between exactness and computational overhead. The accuracy remains sufficiently high even when the actual distribution is different from the modelled distribution. We also conclude that the accuracy is more sensitive to estimation errors related to the \emph{family} of distribution rather than to the \emph{parameters} of the distribution.
(ii) The chance constraints are satisfied almost to the same extent irrespective of the maximum degree~$N_d$. It is much less sensitive to the PCE degree than it is to errors in the estimated distribution.  
}
\vspace{-0.1in}
\subsection{Generation Policies}
\label{sec:GenerationPolicies}
\begin{figure}
	\includegraphics[]{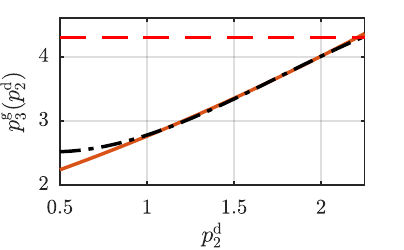}
	\hfill
	\includegraphics[]{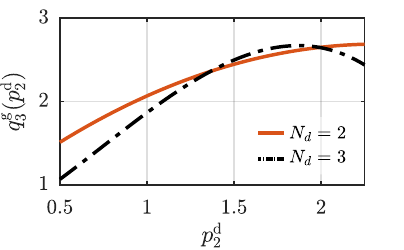}
	\caption{Optimal active/reactive power policies for 5-bus test case.}
	\label{fig:5Bus_ActivePowerPolicies}
	\vspace{\adjustlength}
\end{figure}

{
The solution to the PCE provides a generation control policy which can be evaluated for any realization of uncertainty to provide guidance on how to redispatch generators in a economically efficient and safe manner.
The \pce generator policies for active and reactive power obtained for the 5-bus test case, see Section~\ref{sec:5BusTestCase}, are shown in Figure~\ref{fig:5Bus_ActivePowerPolicies} for maximum degrees $N_d \in \{ 2, 3 \}$.
The modelled uncertainty follows the Beta distribution from Figure~\ref{fig:5Bus_PDF}.}

{We observe that all policies from Figure~\ref{fig:5Bus_ActivePowerPolicies} are non-affine, and that the policies obtained with $N_d=3$ are more curved.
The policies show a significant curvature around $\pd_{2} = 1.1$ for active power and $\pd_{2} = 1.7$, for reactive power, which happens to be the point where an inequality constraint becomes binding. Overall, the reactive power policies have higher curvature than the active power policies owing to stronger nonlinear behavior of reactive power.
We also observe that the upper generation limit $\pplain_{\text{g},3}^{\text{max}} = 4.3$ can be violated by the policy. However, this will be sufficiently unlikely to happen to not exceed the acceptable chance constraint violation probability.}
{
}


	\section{Conclusion and Outlook}
	\label{sec:Conclusions}
	The present paper proposes a tractable reformulation of chance-constrained \acopf using polynomial chaos expansion (PCE).
	PCE allows consideration of the full \ac power flow equations, and it facilitates moment-based reformulations of chance constraints.
	The presented approach requires neither sampling, linearizations nor relaxations.
	The efficacy of the approach is demonstrated for a 5-bus and  a 30-bus system.
Our results indicate that a maximum degree $N_d=2$ for the orthogonal basis provides sufficiently accurate solutions to the \ac power flow under uncertainty, at a manageable computational cost.

	In future work, we will investigate the connection between \pce-based solutions and established control policies such as \textsc{agc}, \textsc{avr}. We would also like to address the question of different cost functions, including e.g., reactive power {or risk-averse minimization of cost variance.}
	{
	To address scalability to larger power grids, we will investigate tailored algorithms to solve the reformulated optimization problem by, e.g., exploiting the sparsity of the nonlinear program.}
{
		Basis-adaptive sparse polynomial chaos should be investigated~\cite{Blatman11}, where, starting from a low-dimensional basis, basis polynomials are added only when they are truly needed.
		}
    Moreover, the effect of adding higher moments to the reformulated chance constraints could provide better reformulations.
    It would also be interesting to investigate how polynomial chaos relates to distributionally robust chance constraints, perhaps merging the advantages of both approaches.
    Finally, N-1 security constraints could be incorporated.
	
\printbibliography
\end{document}